# Comment on
"Spin and spin current - From fundamentals to recent progress"
by Sadamichi Maekawa et al, arXiv: 2211.02241v3

by

Mikhail I. Dyakonov

(University of Montpellier, France;
Ioffe Physical-Technical Institute, Saint Petersburg, Russia)

The first paragraph of this preprint contains the statement: « Electron spins can also flow in a material, independently of their charges. This is called spin current » [references to several recent Maekawa's papers (with co-authors) are provided].

The reader might think that the idea of spin current was introduced in those papers. This would be a wrong conclusion, since the notion of spin current was in fact introduced by Dyakonov and Perel more than 50 years ago (1971). Very strangely, this well known **fact** is never mentioned by Maekawa *et al* in their 40 pages review, which is precisely devoted to spin currents.

The correct statement can be found in many articles, for example, in a review by Yasuhiro Niimi and Yoshi Chika Otani https://arxiv.org/abs/1511.00332: *Reciprocal spin Hall effects in conductors with strong spin-orbit coupling, a review:* « The SHE (spin Hall effect) was originally predicted by Dyakonov and Perel in 1971 and revived by Hirsch about thirty years later. »

The first experimental verification of this prediction in GaAs was obtained in 2004: Kato YK, Myers RC, Gossard AC and Awschalom DD, «Observation of the spin Hall effect in semiconductors» Science **306** 1910 (2004)

In recent years, special attention has been paid to spin current and spin accumulation in semiconductors and metals, and to the problem of utilizing them together with electric charge current. This field of research and applications is named *spintronics*, in contrast with electronics. The flows of spin and charge of electrons can couple to each other due to spin-orbit interactions. As a result, spin current and electric current may convert each to other. The interconversion between the currents is called the spin Hall effect (SHE) and inverse spin Hall effect (ISHE), both predicted theoretically by Dyakonov and Perel in 1971.

See book: *Spin Physics in Semiconductors* (Springer), M.I. Dyakonov editor, 1st edition - 2008, 2nd edition – 2017. See also: https://en.wikipedia.org/wiki/Spin_Hall_effect, as well as the specific effect discovered by M.B. Lifshits and M.I. Dyakonov: Swapping Spin Currents: Interchanging Spin and Flow directions, https://arxiv.org/pdf/0905.4469.pdf